\documentclass[a4paper,11pt]{article}
\usepackage{a4wide,amsmath,amssymb,slashed,bbm}
\usepackage[english]{babel}

\usepackage{tikz}

\usepackage{cite}

\renewcommand{\u}{\underline}
\makeatletter

\@addtoreset{equation}{section} \makeatother

\begin{document}

\hfill{ }

\vspace{30pt}

\begin{center}
{\huge{\bf Integrability of the superstring in \boldmath{$AdS_3\times S^2\times S^2\times T^3$}}}

\vspace{50pt}

Linus Wulff

\vspace{15pt}

{\it\small Department of Theoretical Physics and Astrophysics, Masaryk University, 611 37 Brno, Czech Republic}\\

\vspace{100pt}

{\bf Abstract}
\end{center}
\noindent
Type II supergravity admits an $AdS_3\times S^2\times S^2\times T^3$ solution with fluxes depending on several free parameters. We determine the constraints on these parameters imposed by the requirement of (classical) integrability of the superstring sigma model. To do this we analyze the low-energy effective action for the spinning GKP string. The absence of particle production in the tree-level S-matrix of bosonic excitations is shown to imply the vanishing of two of the four parameters in the NSNS three-form flux. This reduces the supergravity background to either the one-parameter $AdS_3\times S^2\times S^2\times T^3$ background preserving eight supersymmetries, or a non-supersymmetric branch, which differs only by flipping a sign in the RR flux. We show that both these branches can be obtained from $AdS_3\times S^3\times S^3\times S^1$ by T-dualities on the (Hopf) circle fibers of the three-spheres and therefore the integrability of the string in these backgrounds follows.

\pagebreak 
\tableofcontents

\setcounter{page}{1}


\section{Introduction}
Integrability is a very powerful tool in the study of the AdS/CFT correspondence \cite{Maldacena:1997re} since it allows for calculations at any value of the 't Hooft coupling making a precise comparison of the two sides possible, see \cite{Beisert:2010jr} for a review. While the best understood example involves strings in $AdS_5\times S^5$ \cite{Bena:2003wd} there are many other known backgrounds for which the string is (classically) integrable. Most of the examples are symmetric spaces just like $AdS_5\times S^5$. In fact for any symmetric space background preserving some supersymmetry the superstring is classically integrable \cite{Wulff:2015mwa}.\footnote{This was proven to quadratic order in fermions for a generic RR background and for all known backgrounds with both RR and NSNS flux. See also \cite{Wulff:2014kja}.}

Recently a classification of all symmetric space solutions in eleven dimensions was completed \cite{Wulff:2016vqy}, following earlier work \cite{Figueroa-OFarrill:2011tnp}. With such a classification in ten dimensions, which remains to be completed, one can determine the full list of integrable symmetric space AdS backgrounds. Here we take a small step in this direction by analyzing the integrability of strings in the most general\footnote{We are assuming that the fluxes don't break any of the isometries.} $AdS_3\times S^2\times S^2\times T^3$ background. The reason for analyzing this particular background is that it has more free parameters than any other examples we have found. These parameters control the form of the NSNS and RR flux and we want to ask how they are constrained by the requirement of (classical) integrability of the string sigma model.

To do this we focus on a particularly interesting classical string solution, the spinning GKP string \cite{Gubser:2002tv}.\footnote{Specifically we take the limit of a long spinning string for which the coefficients in the string action become constant \cite{Frolov:2002av}. This is also the limit that gives the cusped Wilson loop \cite{Kruczenski:2007cy} and the solution is also closely related to the circular string. In fact another simple way to find the conditions for integrability would be to look at the circular string solution as was done in \cite{Stepanchuk:2012xi} for strings in p-brane backgrounds.} To simplify the analysis further we restrict our attention to the low-energy effective action for the long spinning string obtained by integrating out the massive modes. In the $AdS_5\times S^5$ case this is famously the $O(6)$ sigma model \cite{Alday:2007mf}. In $AdS_4\times\mathbbm{CP}^3$ it is a $\mathbbm{CP}^3$ sigma model coupled to fermions \cite{Bykov:2010tv} while in $AdS_3\times S^3\times S^3\times S^1$ it consists of two $O(4)$ sigma models coupled through non-relativistic quadratic fermion terms \cite{Sundin:2013uca}.

In an integrable theory the S-matrix should factorize into a product of two-particle S-matrices and there should be no particle production \cite{Zamolodchikov:1978xm}. The latter condition requires all two-to-three scattering amplitudes to vanish. By computing some two-to-three tree-level scattering amplitudes of bosons in our low-energy effective model and requiring them to vanish we find that two of the four parameters entering the NSNS flux must vanish. This reduces the background to two one-parameter branches of solutions which differ only in a sign in the RR flux. The first branch preserves eight supersymmetries\footnote{An intersecting brane configuration in type IIB supergravity with this near-horizon geometry was found in \cite{Boonstra:1998yu}. The superisometry group involves the supergroup $D(2,1;\alpha)$ and the Lax connection for the string to quadratic order in fermions was written down in \cite{Wulff:2014kja}.} and the second preserves none. We show that both of these branches can be obtained by starting with $AdS_3\times S^3\times S^3\times S^1$ supported by RR flux and preserving 16 supersymmetries, by performing T-dualities on the circle fibers of the two $S^3$'s viewed as a Hopf fibration over $S^2$ \cite{Duff:1998cr}. The non-supersymmetric branch is obtained by reversing the orientation of one of the $S^3$'s relative to the other. Therefore the string in these backgrounds inherits the integrability from the $AdS_3\times S^3\times S^3\times S^1$ string \cite{Babichenko:2009dk,Sundin:2012gc}. For a detailed analysis in the more general case of squashed $S^3$ we refer to \cite{Orlando:2012hu}.

We demonstrate the duality to $AdS_3\times S^3\times S^3\times S^1$ explicitly for the low-energy GKP string by constructing the action, including fermions, and comparing to the one obtained by T-duality on the Hopf fibers from the model in \cite{Sundin:2013uca}. This gives a nice independent check of the quartic fermion terms in the Green-Schwarz string action derived in \cite{Wulff:2013kga}.
	
The outline of this note is as follows. First we construct the bosonic terms in the low-energy effective GKP string action and show that the absence of particle production constrains the NSNS flux in a crucial way. In section \ref{sec:quadratic} we discuss the supergravity solution and construct the quadratic fermion terms. We then show how to obtain the same action from $AdS_3\times S^3\times S^3\times S^1$ by T-duality in section \ref{sec:Hopf} and end with some conclusions. The quartic fermion terms are worked out in appendix \ref{sec:app}.

\section{Low-energy GKP string: Bosonic terms}
Since we are interested in tree-level scattering of bosonic modes we will drop the fermions in this section. The bosonic string action is
\begin{equation}
S=-\frac{T}{2}\int d^2\xi\left(\gamma^{ij}e_i{}^ae_j{}^b\eta_{ab}-\varepsilon^{ij}B_{ij}\right)\,,\qquad\gamma^{ij}=\sqrt{-g}g^{ij}\,,
\label{eq:Sb}
\end{equation}
where $e_i{}^a$ ($a=0,\ldots,9$) are the pull-backs of the vielbeins and $B_{ij}=e_i{}^ae_j{}^bB_{ab}$ is the pull-back of the NSNS two-form potential. The background we are interested in has geometry $AdS_3\times S^2\times S^2\times T^3$ and is supported by some combination of RR and NSNS flux. Only the NSNS flux is relevant to the bosonic string and it takes the form\footnote{The proof that this is the most general possible form of the NSNS flux will appear elsewhere.}
\begin{equation}
H=dB=2R_{AdS}^{-1}(h_1\sigma_1\wedge dx_7+\sigma_2\wedge(h_2dx_8+h_3dx_7)+h_4dx_7\wedge dx_8\wedge dx_9)\,,
\label{eq:H}
\end{equation}
with four, so far undetermined, dimensionless parameters $h_i$, $i=1,\ldots,4$. $\sigma_{1,2}$ are the volume forms on the two spheres satisfying $\langle\sigma_i,\sigma_j\rangle=\delta_{ij}$ and $x_{7,8,9}$ are the $T^3$ coordinates. The normalization is chosen for later convenience.

To describe the spinning string solution it is convenient to take the $AdS_3$ metric to be \cite{Alday:2007mf,Sundin:2013uca}
\begin{equation}
ds^2=R_{AdS}^2\left(-4dz^+dz^-+4\frac{z+\frac14z^3}{(1-\frac14z^2)^2}\big(-(dz^+)^2+(dz^-)^2\big)+\frac{dz^2}{(1-\frac14z^2)^2}\right)\,.
\label{eq:AdSmetric}
\end{equation}
For the moment we will leave the $S^2$ metric arbitrary.

Following \cite{Sundin:2013uca} we fix conformal gauge $\gamma^{ij}=\eta^{ij}$ and expand around the long spinning string solution
\begin{equation}
z^\pm=\kappa\sigma^\pm\,,\qquad\sigma^\pm=\tfrac12(\tau\pm\sigma)\,.
\label{eq:GKP}
\end{equation}
The next step is to solve the Virasoro constraints for $\partial_\pm z^\mp$ which generates a mass term for $z$. To get the low-energy effective action we then integrate $z$ out. As far as the bosonic action is concerned the effect is simply to remove the terms involving $\partial_+z^-$, $\partial_-z^+$ or $z$. For more details see \cite{Sundin:2013uca}. We are then left with the Lagrangian for two coupled $O(3)$ sigma models plus the three massless bosons coming from the torus directions
\begin{align}
\label{eq:Lbos}
\mathcal L^{\mathrm{bos}}=&
-\frac{T}{2}
\Big(
\eta^{ij}e_i{}^ae_j{}^a
+\eta^{ij}e_i{}^{\hat a}e_j{}^{\hat a}
+\eta^{ij}\partial_ix_7\partial_jx_7
+\eta^{ij}\partial_ix_8\partial_jx_8
+\eta^{ij}\partial_ix_9\partial_jx_9
\\
&{}\qquad
+2h_1R_{AdS}^{-1}x_7\varepsilon^{ij}e_i{}^3e_j{}^4
+2R_{AdS}^{-1}(h_3x_7+h_2x_8)\varepsilon^{ij}e_i{}^5e_j{}^6
+2h_4R_{AdS}^{-1}x_7\varepsilon^{ij}\partial_ix_8\partial_jx_9
\Big)\,,
\nonumber
\end{align}
where $e^a$ $(a=3,4)$ and $e^{\hat a}$ $(\hat a=5,6)$ are the vielbeins of the two $S^2$'s and $x_{7,8,9}$ are the $T^3$ coordinates.

\subsection{Some two-to-three scattering amplitudes}
To compute scattering amplitudes in this model we must pick coordinates for the spheres and expand out the action. Taking stereographic coordinates for the $S^2$'s
\begin{equation}
e^a=R_{S^2_1}\frac{dx^a}{1+\frac14x^mx_m}\,,\qquad (a,m=3,4)
\end{equation}
and similarly for $e^{\hat a}$ we get, upon rescaling the fields so that the kinetic terms are canonically normalized,
\begin{equation}
\mathcal L^{\mathrm{bos}}=\mathcal L_2+\frac{1}{g^{1/2}}\mathcal L_3+\frac{1}{g}\mathcal L_4+\ldots
\end{equation}
with the dimensionless coupling $g=R_{AdS}^2T$.

The quadratic Lagrangian takes the form\footnote{$\partial_\pm=\partial_\tau\pm\partial_\sigma$}
\begin{equation}
\mathcal L_2=
\tfrac12
\left(
\partial_+x_m\partial_-x_m
+\partial_+x_{\hat m}\partial_-x_{\hat m}
+\partial_+u_1\partial_-u_1
+\partial_+u_2\partial_-u_2
+\partial_+u_3\partial_-u_3
\right)\,,
\end{equation}
with $u_{1,2,3}=x_{7,8,9}/R_{AdS}$ dimensionless $T^3$ coordinates. The cubic interaction terms are
\begin{align}
\mathcal L_3=&
\tfrac12h_1u_1(\partial_+x_3\partial_-x_4-\partial_-x_3\partial_+x_4)
+\tfrac12(h_3u_1+h_2u_2)(\partial_+x_5\partial_-x_6-\partial_-x_5\partial_+x_6)
\nonumber\\
&{}
+\tfrac12h_4u_1(\partial_+u_2\partial_-u_3-\partial_-u_2\partial_+u_3)\,,
\label{eq:L3bos}
\end{align}
while the quartic interaction terms are
\begin{equation}
\mathcal L_4=
-\frac{1}{4r_1^2}x_m^2\partial_+x_n\partial_-x_n
-\frac{1}{4r_2^2}x_{\hat m}^2\partial_+x_{\hat n}\partial_-x_{\hat n}
\label{eq:L4bos}
\end{equation}
where $r_1=R_{S^2_1}/R_{AdS}$, $r_2=R_{S^2_2}/R_{AdS}$ are the radii of the spheres in units of the AdS radius, and the quintic ones
\begin{equation}
\mathcal L_5=
-\frac{1}{4r_1^2}h_1u_1x_m^2(\partial_+x_3\partial_-x_4-\partial_-x_3\partial_+x_4)
-\frac{1}{4r_2^2}(h_3u_1+h_2u_2)x_{\hat m}^2(\partial_+x_5\partial_-x_6-\partial_-x_5\partial_+x_6)\,,
\label{eq:L5bos}
\end{equation}
etc.

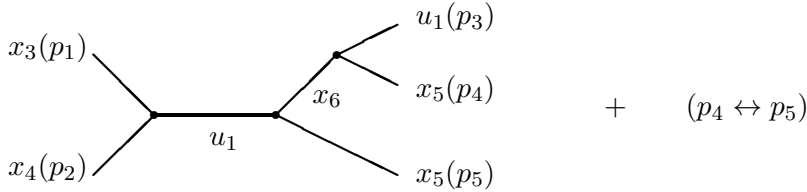
\begin{figure}[ht]
\setlength{\unitlength}{0.8cm}
\begin{picture}(17,4)
\thicklines
\put(2.6,3){$x_3(p_1)$}
\put(2.6,1){$x_4(p_2)$}
\put(4,3){\line(1,-1){1}}
\put(4,1){\line(1,1){1}}
\put(5,2){\circle*{0.15}}
\put(5,2){\line(1,0){2}}
\put(5.9,1.5){$u_1$}
\put(7,2){\circle*{0.15}}
\put(7,2){\line(1,1){1}}
\put(7.6,2.2){$x_6$}
\put(8,3){\line(2,1){1}}
\put(8,3){\line(2,-1){1}}
\put(8,3){\circle*{0.15}}
\put(9.3,3.5){$u_1(p_3)$}
\put(9.3,2.3){$x_5(p_4)$}
\put(7,2){\line(2,-1){2}}
\put(9.3,0.9){$x_5(p_5)$}
\put(12.4,2){$+\qquad(p_4\leftrightarrow p_5)$}
%
\end{picture}
\caption{Tree-level amplitude for $x_3x_4\rightarrow u_1x_5x_5$ scattering.}
\label{fig:x3x4tox5x5u1}
\end{figure}

The requirement of integrability means that there should not be any particle production and therefore that all two-to-three scattering amplitudes should vanish. It will be enough for our purposes to compute the amplitude for $x_3x_4\rightarrow u_1x_5x_5$ and $x_3x_4\rightarrow u_1u_2u_2$ which mix the coordinates $(x_3,x_4,u_1)$ associated with the first $S^2\times S^1$ with $(x_5,x_6,u_2)$ associated with the second $S^2\times S^1$. Looking at the form of the interactions in (\ref{eq:L3bos})--(\ref{eq:L5bos}) it is easy to convince oneself that the only contribution to these tree-level scattering amplitudes come from Feynman diagrams involving only cubic vertices. In fact only one Feynman diagram contributes (modulo exchange of the two identical particles in the final state). The amplitude for $x_3x_4\rightarrow u_1x_5x_5$ is depicted in Figure \ref{fig:x3x4tox5x5u1}.
Evaluating the amplitude in the kinematical regime $p_{1-}=p_{3-}=0$ and $p_{2+}=p_{4+}=p_{5+}=0$ we find
\begin{equation}
\mathcal A(x_3x_4\rightarrow u_1x_5x_5)=i\frac{h_1h_3^2}{8g^{3/2}}p_{1+}p_{2-}\,.
%
%
%
%
\end{equation}
Note that if we replaced $x_5$ with $x_3$ or $x_4$ in the final state there would be additional contributions from other diagrams canceling this. From the form of the interaction terms in (\ref{eq:L3bos})--(\ref{eq:L5bos}) it is clear that the same calculation, but with $h_3\rightarrow h_4$, gives the amplitude
\begin{equation}
\mathcal A(x_3x_4\rightarrow u_1u_2u_2)=i\frac{h_1h_4^2}{8g^{3/2}}p_{1+}p_{2-}\,.
\end{equation}
It is not hard to show that the supergravity equations cannot be satisfied if $H$ only has legs on $T^3$, i.e. if $h_1=h_2=h_3=0$ in (\ref{eq:H}). We can therefore assume, without loss of generality, that $h_1\neq0$. Requiring no particle production then forces $h_3=h_4=0$.

In fact this is enough for the integrability at the bosonic level. To see this we note that with $h_3=h_4=0$ the NSNS three-form $H$ (\ref{eq:H}) satisfies
\begin{equation}
H_{abc'}H^{dec'}\propto R_{ab}{}^{de}\,,\qquad H_{\hat a\hat bc'}H^{\hat d\hat ec'}\propto R_{\hat a\hat b}{}^{\hat c\hat d}
\end{equation}
where the index $c'=7,8,9$ runs over the flat directions and $R_{ab}{}^{cd}$ is the Riemann curvature. In \cite{Wulff:2015mwa} it was shown that a Lax connection exists for the symmetric space bosonic string if this condition is satisfied. Therefore there are no further conditions needed for the integrability of the bosonic string. However, the superstring action contains also fermionic terms, and we now turn to a discussion of these.

\section{Quadratic fermion terms}\label{sec:quadratic}
To write the terms in the string action involving fermions we first of all need to know the full supergravity solution including the RR fields. Taking the most general possible ansatz for the RR fluxes compatible with the isometries of $AdS_3\times S^2\times S^2\times T^3$ and the form of $H$ in (\ref{eq:H}) with $h_3=h_4=0$ we find the type IIA supergravity solution\footnote{It is of course trivial to T-dualize this solution along the $x_9$-direction to obtain the corresponding type IIB solution.}
\begin{align}
H=&2R_{AdS}^{-1}(h_1\sigma_1\wedge dx_7+h_2\sigma_2\wedge dx_8)\,,\label{eq:H2}\\
F^{(4)}=&2R_{AdS}^{-1}(\sigma_1\wedge\sigma_2+h_1\sigma_1\wedge dx_8\wedge dx_9\pm h_2\sigma_2\wedge dx_7\wedge dx_9)\,,
\end{align}
with $h_1,h_2$ determining the radii of the two $S^2$'s as $r_1=1/(2h_1)$ and $r_2=1/(2h_2)$ or, in our conventions $R_{34}{}^{34}=-(2h_1/R_{AdS})^2$ and $R_{56}{}^{56}=-(2h_2/R_{AdS})^2$, and satisfying the constraint
\begin{equation}
h_1^2+h_2^2=1\,.
\end{equation}
Another way to obtain this solution is by dimensional reduction of the corresponding $D=11$ solution in \cite{Wulff:2016vqy}. Note that there are two branches of the solution differing only in the sign of the last term in the RR four-form flux. The branch with the upper sign is non-supersymmetric while the branch with the lower sign preserves eight supersymmetries.\footnote{Supersymmetric (warped) $AdS_3\times S^2\times\mathcal M_6$ solutions in $D=11$, with $\mathcal M_6$ an $SU(2)$-structure manifold, were classified in \cite{Kelekci:2016uqv}. This includes the lift of the supersymmetric $AdS_3\times S^2\times S^2\times T^3$ solution originally found in \cite{Boonstra:1998yu}.} To see this we look at the supersymmetry conditions, which consist of the dilatino equation and the integrability condition for the Killing spinor equation,
\begin{equation}
\mathcal T\xi=0\,,\qquad U_{ab}\xi=0\,,
\end{equation}
where for constant fluxes and dilaton we have (e.g. \cite{Wulff:2013kga})
\begin{equation}
\mathcal T=\tfrac{i}{24}H_{abc}\Gamma^{abc}\Gamma_{11}+\tfrac{i}{16}\Gamma^a\mathcal S\Gamma_a\,,\qquad U_{ab}=\tfrac{1}{32}G_{[a}G_{b]}-\tfrac14R_{ab}{}^{cd}\Gamma_{cd}\,,\qquad G_a=H_{abc}\Gamma^{bc}\Gamma_{11}+\mathcal S\Gamma_a\,,
\label{eq:TandUab}
\end{equation}
where $\mathcal S$ is the RR bispinor which in our case is $\mathcal S=\slashed F^{(4)}$. Explicitly we have
\begin{equation}
\mathcal S=4R_{AdS}^{-1}\Gamma^{3456}P_2\,,\qquad
\mathcal T=\frac{i}{2R_{AdS}}\Gamma^{3456}\left[P_2+\Gamma^{789}\Gamma_{11}(h_1\Gamma^{5689}-h_2\Gamma^{3479})\right]
\label{eq:SandT}
\end{equation}
where we have defined the projection operator
\begin{equation}
P_2=\tfrac12(1-h_1\Gamma^{5689}\mp h_2\Gamma^{3479})
\label{eq:P2}
\end{equation}
which projects on half of the fermions. The supersymmetry condition $0=U_{09}\xi=\tfrac{1}{32}\mathcal S\Gamma_{[0}\mathcal S\Gamma_{9]}\xi$ implies the projection condition on the Killing spinor $(1-P_2)\xi=0$. The dilatino equation then says
\begin{equation}
\left(
1-\Gamma^{789}\Gamma_{11}
-(1\pm1)h_2\Gamma^{3479}\Gamma^{789}\Gamma_{11}
\right)\xi=0
\end{equation}
so that for the lower sign $(1-P_1)\xi=0$ with $P_1=\frac12(1+\Gamma^{789}\Gamma_{11})$, in which case also the other supersymmetry conditions are satisfied. For the upper sign one finds, by squaring the operator acting on $\xi$ and using the equation again, that $h_2(h_2+\Gamma^{3479})\xi=0$ which has no solution unless $h_2=0,1$ in which case it is equivalent to the supersymmetric branch.

We are now ready to write down the fermionic terms in the low-energy effective GKP string action. On dimensional grounds it can contain at most quartic fermion terms so that
\begin{equation}
\mathcal L=\mathcal L^{\mathrm{bos}}+\mathcal L^{\mathrm{2ferm}}+\mathcal L^{\mathrm{4ferm}}\,.
\label{eq:L0+L2+L4}
\end{equation}
We will defer the discussion of the quartic fermion terms to Appendix \ref{sec:app}. To derive the quadratic fermion terms we start from the quadratic terms in the type IIA Green-Schwarz string action \cite{Wulff:2013kga}\footnote{Here $\theta$ is a 32-component Majorana fermion and we suppress the charge conjugation matrix, i.e. $\theta\Gamma^a\kappa=\theta^\alpha(C\Gamma^a)_{\alpha\beta}\kappa^\beta$.}
\begin{equation}
-iTe_i{}^a\,\theta\Gamma_a(\gamma^{ij}-\varepsilon^{ij}\Gamma_{11})\mathcal D_j\theta\,,
\label{eq:L2-GS}
\end{equation}
where $\mathcal D_j=\partial_j-\frac14\omega_j{}^{ab}\Gamma_{ab}+\frac18e_j{}^a(H_{abc}\Gamma^{bc}\Gamma_{11}+\mathcal S\Gamma_a)$. Plugging in the GKP string solution (\ref{eq:GKP}) one finds that the kinetic term for the fermions involves the projector
\begin{equation}
P=\tfrac12(1-\Gamma^{01}\Gamma_{11})\,.
\end{equation}
This is a manifestation of the kappa symmetry of the Green-Schwarz string action which guarantees that only 16 of the 32 fermions are physical. We fix it by imposing the natural kappa symmetry gauge
\begin{equation}
\theta=P\theta\,.
\label{eq:kappa-gauge}
\end{equation}
Doing this the quadratic terms in the Green-Schwarz action involving the fermions take the form (rescaling the fermions appropriately)
\begin{equation}
\mathcal L_2^{\mathrm{2ferm}}=
-2i\theta\Gamma^0\partial_0\theta
-2i\theta\Gamma^1\partial_1\theta
-2i\theta\Gamma^{3456}P_2\theta
\end{equation}
where we used the form of $\mathcal S$ in (\ref{eq:SandT}). We see that the fermions $P_2\theta$ are massive and we should drop them in the low-energy effective action and leave only the massless ones
\begin{equation}
\tilde\upsilon=(1-P_2)P\theta\,,
\end{equation}
where the second projector comes from the kappa symmetry gauge fixing. Using the expansion of the AdS vielbein and spin connection in the coordinates (\ref{eq:AdSmetric})
\begin{equation}
e^\pm=R_{AdS}(dz^\pm\mp zdz^\mp)+\ldots
\,,\qquad\omega^{2\pm}=\mp dz^{\mp}+zdz^\pm+\ldots
\label{eq:ads3-vielbeins}
\end{equation}
and dropping terms of mass dimension higher than two we find for the low-energy effective action
\begin{align}
\mathcal L^{\mathrm{2ferm}}=&
-2iTR_{AdS}\tilde\upsilon\Gamma^i(\partial_i-\tfrac12\omega_i{}^{34}\Gamma_{34}-\tfrac12\omega_i{}^{56}\Gamma_{56})\tilde\upsilon
-iT\delta^{ij}e_i{}^{a'}\,\tilde\upsilon\Gamma_{a'2}\Gamma_j\Gamma_{11}\tilde\upsilon
\nonumber\\
&{}
-\tfrac{i}{4}TR_{AdS}e_i{}^{a'}H_{a'b'c'}\,\tilde\upsilon\Gamma^i\Gamma^{b'c'}\Gamma_{11}\tilde\upsilon\,,
\end{align}
where $i,j=0,1$ and $a'=3,\ldots,9$ runs over all non-AdS directions. Note that the second term, which comes from the AdS spin connection, breaks 2d Lorentz invariance as in \cite{Sundin:2013uca}. Using the form of $H_{abc}$ in (\ref{eq:H2}) and $P_2$ in (\ref{eq:P2}) and expressing the eight fermions in terms of four dimensionless 2d Majorana fermions 
\begin{equation}
\tilde\upsilon\rightarrow i\frac{R_{AdS}^{1/2}}{2}\psi_{\pm\pm}
\label{eq:psi}
\end{equation}
on which the $32\times32$ gamma-matrices and charge-conjugation matrix act as
\begin{align}
&C\rightarrow\rho^0\otimes\sigma^2\otimes\sigma^2\,,\quad
\Gamma^i\rightarrow\rho^i\otimes\sigma^2\otimes\sigma^2\,,\quad
\Gamma^{34}\rightarrow i\mathbbm1\otimes T_3\,,\quad
\Gamma^{56}\rightarrow i\mathbbm1\otimes\hat T_3\,,\quad
\nonumber\\
&\Gamma^{37}\rightarrow i\rho^3\otimes T_2\,,\quad
\Gamma^{58}\rightarrow i\rho^3\otimes\hat T_2\,,
\label{eq:gammas}
\end{align}
with 2d gamma-matrices
\begin{equation}
\rho^0=i\sigma^2\,,\qquad\rho^1=\sigma^1\,,\qquad\rho^3=\rho^0\rho^1
\end{equation}
and two sets of $SU(2)$ generators
\begin{equation}
T_1=\sigma^3\otimes\sigma^2\,,\quad
T_2=\sigma^1\otimes\sigma^2\,,\quad
T_3=\sigma^2\otimes\mathbbm1\,,\qquad
\hat T_1=\sigma^2\otimes\sigma^3\,,\quad
\hat T_2=\sigma^2\otimes\sigma^1\,,\quad
\hat T_3=\mathbbm1\otimes\sigma^2\,,
\label{eq:Ts}
\end{equation}
the quadratic fermion terms in the low-energy effective GKP string action take the simple 2d form\footnote{Explicitly $\psi\rho^iT_a\psi=\psi^\alpha_I(\rho^0\rho^i)_{\alpha\beta}(T_a)_{IJ}\psi^\beta_J$ with $I,J=++,+-,-+,--$.}
\begin{align}
g^{-1}\mathcal L^{\mathrm{2ferm}}=&
%
%
\tfrac{i}{2}\psi\rho^i(\partial_i-\tfrac{i}{2}\omega_i{}^{34}T_3-\tfrac{i}{2}\omega_i{}^{56}\hat T_3)\psi
-\tfrac14h_1R_{AdS}^{-1}(\eta^{ij}+\delta^{ij})e_i{}^a\,\psi\rho_jT_a\psi
\nonumber\\
&{}
+\tfrac14h_1(\eta^{ij}+\delta^{ij})\partial_iu_1\,\psi\rho_j\rho^3T_3\psi
-\tfrac14h_2R_{AdS}^{-1}(\eta^{ij}\mp\delta^{ij})e_i{}^{\hat a}\,\psi\rho_j\hat T_{\hat a}\psi
\nonumber\\
&{}
+\tfrac14h_2(\eta^{ij}\mp\delta^{ij})\partial_iu_2\,\psi\rho_j\rho^3\hat T_3\psi\,.
\label{eq:L2ferm}
\end{align}
Note that the last torus coordinate $u_3=x_9/R_{AdS}$ decouples. Together with the bosonic terms in (\ref{eq:Lbos}) and the quartic fermion terms derived in Appendix \ref{sec:app} this gives the complete low-energy effective action for the $AdS_3\times S^2\times S^2\times T^3$ GKP string. In the next section we will show that the same action can be obtained from the low-energy effective action for the $AdS_3\times S^3\times S^3\times S^1$ GKP string written down in \cite{Sundin:2013uca} by two T-dualities on the Hopf fibers of the three-spheres. This also proves the classical integrability of the model.

\section{T-duality from \boldmath{$AdS_3\times S^3\times S^3\times S^1$}}\label{sec:Hopf}
One can show that the one-parameter family of $AdS_3\times S^2\times S^2\times T^3$ supergravity backgrounds with fluxes given in (\ref{eq:H2}) can be obtained from the $AdS_3\times S^3\times S^3\times S^1$ type IIA supergravity solution supported by RR four-form flux and preserving 16 supersymmetries by viewing $S^3$ as an $S^1$-fibration over $\mathbbm{CP}^1\simeq S^2$ and T-dualizing on the circle fibers.\footnote{For a detailed analysis in the supersymmetric case see appendix A of \cite{Lozano:2015bra}.} Here we will instead demonstrate explicitly how this works at the level of the low-energy GKP string sigma model. For simplicity we consider only one of the $S^3$-factors in the low-energy effective GKP string in $AdS_3\times S^3\times S^3\times S^1$. The Lagrangian is \cite{Sundin:2013uca}\footnote{In that reference a slightly different notation was used for the fermions representing them instead as $2\times2$-matrices.}
\begin{equation}
\mathcal L_{S^3}=\tfrac12\eta^{ij}e_i{}^{\u a}e_j{}^{\u a}+\tfrac{i}{2}\psi\rho^i(\partial_i-\tfrac{i}{4}\u\omega_i{}^{\u a\u b}\varepsilon_{\u a\u b\u c}T_{\u c})\psi-\tfrac14\delta^{ij}e_i{}^{\u a}\psi\rho_jT_{\u a}\psi\,,
\label{eq:S3-model}
\end{equation}
with $S^3$ indices $\u a=1,2,3$ and $T_{\u a}$ given in (\ref{eq:Ts}). Next we take a frame adapted to the Hopf fibration structure of $S^3$ (we set the $S^3$ radius to 1)
\begin{equation}
e^{\u a}=(e^a,e^3)\,,\qquad e^3=dy+dx^mA_m(x)\,,
\end{equation}
$y$ denoting the fiber coordinate and $x^m$ ($m=1,2$) coordinates of the base. Similarly the spin connection takes the form
\begin{equation}
\u\omega^{ab}=\omega^{ab}+\varepsilon^{ab}e^3\,,\qquad\omega^{a3}=\varepsilon^{ab}e_b\,,
\end{equation}
and the curvature of the gauge field is given by $dA=e^be^a\varepsilon_{ab}$. This ansatz describes $S^3$ as a Hopf fibration over $S^2$ (of radius $\frac12$) and plugging into the Lagrangian (\ref{eq:S3-model}) we get
\begin{align}
\mathcal L'_{S^3}=&
\tfrac12\eta^{ij}e_i{}^ae_j{}^a
+\tfrac{i}{2}\psi\rho^i(\partial_i-\tfrac{i}{2}\hat\omega_i{}^{12}T_3)\psi
-\tfrac14(\eta^{ij}+\delta^{ij})e_i{}^a\,\psi\rho_jT_a\psi
+\tfrac12(\partial_iy+A_i)(\partial^iy+A^i)
\nonumber\\
&{}
+\tfrac14(\eta^{ij}-\delta^{ij})(\partial_iy+A_i)\,\psi\rho_jT_3\psi
-\tilde y\varepsilon^{ij}(\partial_iA_j-e_i{}^ae_j{}^b\varepsilon_{ab})\,,
\end{align}
where we have included a Lagrange multiplier $\tilde y$ enforcing the constraint on the curvature of $A$. Integrating out $\tilde y$ gives the original model while integrating out $A$ instead gives the T-dual model. We find
\begin{equation}
\partial^iy+A^i
=
\varepsilon^{ij}\partial_j\tilde y
-\tfrac14(\eta^{ij}-\delta^{ij})\,\psi\rho_jT_3\psi
\end{equation}
and using this the dualized Lagrangian becomes
\begin{align}
\label{eq:L-dual}
\tilde{\mathcal L}_{S^2\times S^1}=&
\tfrac12\eta^{ij}e_i{}^ae_j{}^a
+\tfrac12\partial_i\tilde y\partial^i\tilde y
+\tfrac12\tilde y\varepsilon^{ij}e_i{}^ae_j{}^b\varepsilon_{ab}
+\tfrac{i}{2}\psi\rho^i(\partial_i-\frac{i}{2}\hat\omega_i{}^{12}T_3)\psi
\\
&{}
-\tfrac14(\eta^{ij}+\delta^{ij})e_i{}^a\,\psi\rho_jT_a\psi
+\tfrac14(\eta^{ij}+\delta^{ij})\partial_i\tilde y\,\psi\rho_j\rho^3T_3\psi
-\tfrac{1}{16}(\eta^{ij}-\delta^{ij})\psi\rho_iT_3\psi\,\psi\rho_jT_3\psi\,.
\nonumber
\end{align}
Comparing this to the low-energy effective action for the string in $AdS_3\times S^2\times S^2\times T^3$ in (\ref{eq:Lbos}) and (\ref{eq:L2ferm}) we see that they match precisely when $h_1=1$, $h_2=0$. Notice further that if we reverse the orientation of the $S^3$ by changing the sign of $e^{\u a}$ in (\ref{eq:S3-model}) all steps go through as above but with the sign of $\delta^{ij}$ changed everywhere. But this is precisely what we have in (\ref{eq:Lbos}) and (\ref{eq:L2ferm}) for the terms involving the other $S^2$ factor, i.e. when $h_1=0$, $h_2=1$. It is therefore clear that if one starts with the general low-energy effective GKP action in $AdS_3\times S^3\times S^3\times S^1$ which consists of two factors of the form (\ref{eq:S3-model}) \cite{Sundin:2013uca}, with the $S^3$ radii being $R_{AdS}/h_1$ and $R_{AdS}/h_2$, respectively, one obtains, after T-duality on the Hopf fibers of the two $S^3$'s precisely the low-energy effective action for the string in $AdS_3\times S^2\times S^2\times T^3$ in (\ref{eq:Lbos}) and (\ref{eq:L2ferm}). The supersymmetric/non-supersymmetric branch corresponding to taking the same/opposite orientation of the $S^3$'s.

A Lax connection demonstrating the integrability of the model can be easily found by implementing the T-duality in the Lax connection for the $AdS_3\times S^3\times S^3\times S^1$ model found in \cite{Sundin:2013uca}.

\section{Conclusions}
We have shown that the most general symmetric space $AdS_3\times S^2\times S^2\times T^3$ background for which the superstring is classically integrable is one of the two branches which can be obtained from $AdS_3\times S^3\times S^3\times S^1$ supported by RR flux by T-duality. This suggests that supersymmetry is important for integrability since, even though one branch is non-supersymmetric at the supergravity level, the supersymmetry is still there in the full string theory through the duality to $AdS_3\times S^3\times S^3\times S^1$. This is consistent with the important role played by supersymmetry in constructing Lax connections in \cite{Wulff:2014kja,Wulff:2015mwa}. Note that in a similar way one can obtain more general integrable backgrounds involving warped $AdS_3$ and/or squashed $S^3$'s by starting instead from $AdS_3\times S^3\times S^3\times S^1$ supported by a mix of RR and NSNS flux and performing T-dualities on the Hopf fibers of the three-spheres \cite{Duff:1998cr,Orlando:2010ay}.\footnote{The integrability of the $AdS_3\times S^3\times S^3\times S^1$ string with mixed flux was analyzed in \cite{Cagnazzo:2012se,Wulff:2014kja}.} Integrability of the string in this general setting was analyzed in \cite{Orlando:2010yh,Orlando:2012hu}. A special case gives integrable $AdS_2$ backgrounds, many of which were analyzed in \cite{Wulff:2014kja,Wulff:2015mwa}.

Applying similar arguments to the ones considered here to other backgrounds it should be possible to complete the classification of integrable symmetric space backgrounds.\footnote{For $AdS_2$ backgrounds there is no GKP string and one should use another classical string solution, e.g. the circular string used in \cite{Stepanchuk:2012xi}.} This would also help answer the question of whether integrability of the bosonic string guarantees integrability of the superstring. Despite the presence of kappa symmetry, which relates the bosons and fermions, I believe the answer to be no unless the background preserves some supersymmetry \cite{Wulff:2014kja}. I hope to report on this in the near future.

\vspace{1cm}

\section*{Acknowledgments}
I wish to thank A. Tseytlin for comments on a draft of this note.

\vspace{0.5cm}

\noindent{\LARGE\bf Appendix}

\appendix

\section{Quartic fermion terms}\label{sec:app}
Quartic fermion terms in the low-energy effective action can be generated by couplings of the form $\partial_-z^+\theta^2$ or $\partial_+z^-\theta^2$ upon solving the Virasoro constraints for $\partial_\pm z^\mp$. They can also be generated by couplings of the form $z\theta^2$ upon integrating out the massive AdS boson $z$. Expanding out the Lagrangian (\ref{eq:L2-GS}) one finds that the relevant terms are\footnote{It is easy to see that only the AdS vielbein and spin connection terms contribute, and the result then follows using (\ref{eq:ads3-vielbeins}).}
\begin{equation}
\tfrac{i}{2}TR_{AdS}(\partial_-z^++\partial_+z^-)\,\theta\Gamma_{2+-}\theta
-iTR_{AdS}z\,\theta\Gamma_{2+-}\Gamma_{11}\theta\,.
\end{equation}
However, these terms vanish when we impose the kappa symmetry gauge fixing (\ref{eq:kappa-gauge}). Therefore no $\theta^4$-terms are generated by solving the Virasoro constraints or integrating out the massive AdS boson $z$.

All $\theta^4$-terms therefore come from the corresponding part of the Green-Schwarz superstring action. These were determined for a general supergravity background in \cite{Wulff:2013kga} and in the type IIA case take the form\footnote{The fermions have been rescaled by a factor of $\sqrt2$ compared to \cite{Wulff:2013kga} to be consistent with the normalization of the $\theta^2$-terms used here.}
\begin{align}
T^{-1}\mathcal L^{(4)}=&
-\tfrac12\theta\Gamma^a\mathcal D_i\theta\,\theta\Gamma_a(\gamma^{ij}-\varepsilon^{ij}\Gamma_{11})\mathcal D_j\theta
+\tfrac{i}{6}e_i{}^a\,\theta\Gamma_a(\gamma^{ij}-\varepsilon^{ij}\Gamma_{11})\mathcal M\mathcal D_j\theta
\nonumber\\
&{}
+\tfrac{i}{48}e_i{}^ae_j{}^b\,\theta\Gamma_a(\gamma^{ij}-\varepsilon^{ij}\Gamma_{11})(M+\tilde M)\mathcal S\Gamma_b\theta
\nonumber\\
&{}
+\tfrac{1}{48}e_i{}^ce_j{}^d\,\theta\Gamma_c{}^{ab}(\gamma^{ij}-\varepsilon^{ij}\Gamma_{11})\theta\,\big[3\theta\Gamma_dU_{ab}\theta-2\theta\Gamma_aU_{bd}\theta\big]
\nonumber\\
&{}
-\tfrac{1}{48}e_i{}^ce_j{}^d\,\theta\Gamma_c{}^{ab}\Gamma_{11}(\gamma^{ij}-\varepsilon^{ij}\Gamma_{11})\theta\,\big[3\theta\Gamma_d\Gamma_{11}U_{ab}\theta+2\theta\Gamma_a\Gamma_{11}U_{bd}\theta\big]
\label{eq:L4-GS}
\end{align}
where
\begin{align}
\mathcal M=&M+\tilde M
+\tfrac{i}{8}(G_a\theta)(\theta\Gamma^a)
-\tfrac{i}{16}(\Gamma^{ab}\theta)(\theta[\Gamma_a\mathcal S\Gamma_b-2H_{abc}\Gamma^c\Gamma_{11}])
\label{eq:M}
\\
M=&\tfrac12\theta\mathcal T\theta\cdot1-\tfrac12\theta\Gamma_{11}\mathcal T\theta\cdot\Gamma_{11}+(\theta)(\mathcal T\theta)+(\Gamma^a\mathcal T\theta)(\theta\Gamma_a)\,,\qquad\tilde M=\Gamma_{11}M\Gamma_{11}\,.
\end{align}
For constant fluxes and dilaton $\mathcal T$ and $U_{ab}$ are given by (\ref{eq:TandUab}).

We now evaluate the contribution from (\ref{eq:L4-GS}) to the low-energy effective GKP string action. From the third term we get, using (\ref{eq:SandT}), terms involving the massive fermions which give no contribution to the low-energy effective action. The only contribution to the low-energy effective action from the first two terms comes from the AdS spin connection inside $\mathcal D$. The only contribution from the second term comes from the terms in $\mathcal M$, (\ref{eq:M}), involving $H_{abc}$ and from the last term in $M$($\tilde M$) and we find, using the kappa projection (\ref{eq:kappa-gauge}),\footnote{$\Gamma_\pm=\Gamma_0\pm\Gamma_1$.}
\begin{align}
T^{-1}\mathcal L^{4\mathrm{ferm}}=&
-\tfrac14\tilde\upsilon\Gamma_{2+}{}^{a'}\tilde\upsilon\,\tilde\upsilon\Gamma_{2-a'}\tilde\upsilon
-\tfrac{iR_{AdS}}{6}\tilde\upsilon\Gamma_+\Gamma^{a'}\mathcal T\tilde\upsilon\,\tilde\upsilon\Gamma_{2+a'}\tilde\upsilon
+\tfrac{iR_{AdS}}{6}\tilde\upsilon\Gamma_-\Gamma^{a'}\mathcal T\tilde\upsilon\,\tilde\upsilon\Gamma_{2-a'}\tilde\upsilon
\nonumber\\
&{}
-\tfrac{R_{AdS}}{48}H^{a'b'c'}\tilde\upsilon\Gamma_{+a'b'}\tilde\upsilon\,\tilde\upsilon\Gamma_{2+c'}\tilde\upsilon
-\tfrac{R_{AdS}}{48}H^{a'b'c'}\tilde\upsilon\Gamma_{-a'b'}\tilde\upsilon\,\tilde\upsilon\Gamma_{2-c'}\tilde\upsilon
\nonumber\\
&{}
-\tfrac{R_{AdS}^2}{24}\tilde\upsilon\Gamma_+{}^{ab}\tilde\upsilon\,\big[3\tilde\upsilon\Gamma_-U_{ab}\tilde\upsilon-\tilde\upsilon\Gamma_a(1-\Gamma_{11})U_{b-}\tilde\upsilon\big]
\nonumber\\
&{}
-\tfrac{R_{AdS}^2}{24}\tilde\upsilon\Gamma_-{}^{ab}\tilde\upsilon\,\big[3\tilde\upsilon\Gamma_+U_{ab}\tilde\upsilon-\tilde\upsilon\Gamma_a(1+\Gamma_{11})U_{b+}\tilde\upsilon\big]\,.
\end{align}
From the form of $U_{ab}$, (\ref{eq:TandUab}), it is easy to see that the terms with $U_{b\pm}$ give nothing and the only contribution from the $U_{ab}$ terms comes from $H_{abc}$ and the Riemann tensor. Furthermore, using the form of the projector $P_2$, (\ref{eq:P2}), one can show that the $H_{abc}$ terms from $U_{ab}$ cancel against the first term. Using (\ref{eq:H2}) and (\ref{eq:SandT}) the remaining terms become
\begin{align}
T^{-1}\mathcal L^{4\mathrm{ferm}}=&
-\tfrac{h_1^2}{6}(\tilde\upsilon\Gamma_{+a7}\tilde\upsilon)^2
-\tfrac{h_1^2}{6}(\tilde\upsilon\Gamma_{-a7}\tilde\upsilon)^2
-\tfrac{h_1^2}{6}(\tilde\upsilon\Gamma_{+34}\tilde\upsilon)^2
-\tfrac{h_1^2}{6}(\tilde\upsilon\Gamma_{-34}\tilde\upsilon)^2
\nonumber\\
&{}
\pm\tfrac{h_2^2}{6}(\tilde\upsilon\Gamma_{+\hat a8}\tilde\upsilon)^2
\pm\tfrac{h_2^2}{6}(\tilde\upsilon\Gamma_{-\hat a8}\tilde\upsilon)^2
\pm\tfrac{h_2^2}{6}(\tilde\upsilon\Gamma_{+56}\tilde\upsilon)^2
\pm\tfrac{h_2^2}{6}(\tilde\upsilon\Gamma_{-56}\tilde\upsilon)^2
\nonumber\\
&{}
-h_1^2\tilde\upsilon\Gamma_{+34}\tilde\upsilon\,\tilde\upsilon\Gamma_{-34}\tilde\upsilon
-h_2^2\tilde\upsilon\Gamma_{+56}\tilde\upsilon\,\tilde\upsilon\Gamma_{-56}\tilde\upsilon\,.
\end{align}
Finally, passing to 2d notation via (\ref{eq:psi}) and (\ref{eq:gammas}), and noting that due to the Grassmann nature of $\psi$ we have
\begin{align}
(\psi\rho_+T_a\psi)^2
=&
-4(
\psi_{++}\rho_+\psi_{--}
-\psi_{+-}\rho_+\psi_{-+}
)^2
-4(
\psi_{++}\rho_+\psi_{+-}
-\psi_{-+}\rho_+\psi_{--}
)^2
\nonumber\\
=&
16\psi_{++}\rho_+\psi_{--}\,\psi_{+-}\rho_+\psi_{-+}
=
2(\psi\rho_+T_3\psi)^2
\end{align}
and similarly for $\hat T_a$ defined in (\ref{eq:Ts}), we find
\begin{align}
g^{-1}\mathcal L^{4\mathrm{ferm}}=
-\tfrac{h_1^2}{16}(\eta^{ij}-\delta^{ij})\psi\rho_iT_3\psi\,\psi\rho_jT_3\psi
-\tfrac{h_2^2}{16}(\eta^{ij}\pm\delta^{ij})\psi\rho_iT_3\psi\,\psi\rho_jT_3\psi\,.
\end{align}
This expression matches precisely the quartic fermion terms which one obtains by T-duality from $AdS_3\times S^3\times S^3\times S^1$, cf. (\ref{eq:L-dual}).

\end{document}